\documentclass[12pt,final,english]{iopart}
\usepackage[T1]{fontenc}
\usepackage[latin1]{inputenc}
\usepackage{graphicx}
\usepackage{setspace}
\usepackage{color}
\makeatletter

\newcommand {\bb}           {\mathbf{b}}
\newcommand {\bB}           {\mathbf{B}}
\newcommand {\bk}           {\mathbf{k}}

\newcommand{\mody}{\color{black}}
\newcommand{\modb}{\color{black}}
\newcommand{\modc}{\color{black}}
\newcommand{\norm}{\color{black}}

\usepackage{babel}
\makeatother
\begin{document}

\title[Beta-induced Alfv\'{e}n wave activity in KSTAR plasmas]
{First evidence of Alfv\'{e}n wave activity in KSTAR plasmas}

\author{M.J. Hole$^1$, C. M. Ryu$^2$, M.H. Woo$^2$, J. G. Bak$^3$, S E. Sharapov$^4$, M. Fitzgerald$^1$, and the KSTAR team$^3$}

\address{$^1$ Research School of Physical Sciences and Engineering, Australian National University, Acton 0200, ACT Australia}
\address{$^2$ POSTECH, Pohang, Korea}
\address{$^3$ National Fusion Research Institute, Daejeon, Korea}
\address{$^4$ EURATOM/CCFE Fusion Assoc., Culham Science Centre,Abingdon, Oxon OX14 3DB, UK}

\begin{abstract}
We report on first evidence of wave activity \mody during neutral beam heating \norm in KSTAR plasmas: 
40~kHz magnetic fluctuations with a toroidal mode number of $n=1$.
Our analysis suggests this a beta-induced Alfv\'{e}n eigenmode resonant with the $q=1$ surface.
A  kinetic analysis, when coupled with electron temperature measurements from electron cyclotron emission and ion/electron temperature ratios from crystallography, 
enables calculation of the frequency evolution, which is in agreement with observations.  
Complementary detailed MHD modelling of the magnetic configuration and wave modes supports the BAE mode conclusion, by locating an $n=1$ mode
separated from the continuum in the core region.  
Finally, we have computed the threshold to marginal stability for a range of ion temperature profiles. These suggest the BAE can be driven unstable \modc by 
energetic ions \norm
when the ion temperature radial gradient is sufficiently large.
Our findings suggest that mode existence could be used as a form of inference for 
temperature profile consistency in the radial interval of the mode, thereby extending the tools of MHD spectroscopy.

\end{abstract}

% Uncomment for PACS numbers title message
% 52.55.Fa Tokamaks, spherical tokamaks
% 52.55.-s Magnetic confinement and equilibrium (see also 28.52.-s Fusion reactors)
% 52.55.Tn Ideal and resistive MHD modes; kinetic modes
% 28.52.-s Fusion reactors (see also 52.55.-s, 52.57.-z, and 52.58.-c in physics of plasmas)
\pacs{52.55.Fa, 52.55.-s,52.55.Tn}
 
\submitto{\PPCF}

% Comment nout if separate title page not required
\maketitle

% Include Table of contents - command must be wrong
% \tableofcontents

% #################################################################################################
\section{Introduction}

Instabilities such as Alfv\'{e}n eigenmodes, driven by fast particles, are of programmatic concern as they can 
expel energetic ions from the plasma, thereby preventing heating by thermalisation. \cite{Breizman_11} In addition, such 
energetic particles expelled can damage the first wall, and a fusion reactor can only tolerate fast particle losses of a few per cent. \cite{Pinches_04}
Another motivation for the study of Alfv\'{e}n eigenmodes is their potential use as a diagnostic for the plasma, 
particularly through the tool of MHD spectroscopy. \cite{Holties_97} 

One such class of fast particle driven instabilities that can occur at relatively low frequency are beta induced Alfv\'{e}n eigenmodes (BAEs). 
The characteristic experimental feature of this instability are magnetic fluctuations at a frequency intermediate between 
the fishbone and the toroidal Alfv\'{e}n eigenmode (TAE), with angular frequency $\omega_{\mathrm{TAE}} = v_A/(2 q R)$, with $v_A$ the Alfv\'{e}n speed, 
$q$ the safety factor and $R$ the major radius. \cite{Heidbrink_99}
These modes were first identified in DIII-D, \cite{Heidbrink_93} and have since been discovered in other tokamaks in beam, \cite{Nazikian_96} 
ion cyclotron heated \cite{Nguyen_09} and Ohmically heated discharges in the presence of a magnetic island. \cite{Smeulders_02, Buratti_05a, Buratti_05b, Zimmerman_05, Annibaldi_07} 
\mody More advanced kinetic treatments, which include corrections to kinetic theory for diamagnetic and shaping effects, as well as the 
inclusion of trapped particles have been used to study BAEs in ASDEX. \cite{Lauber_09} Radial profile information was measured during sawteeth in Tore Supra. \cite{Guimaraes_11}  \norm
Recently, BAEs driven by electron populations have been observed in the tokamak HL-2A, \cite{Chen_BAE_10} and there is some evidence that
magnetic oscillations in the H-1 heliac are also driven by energetic electrons. \cite{Bertram_12} 

The aim of this paper is to report on first evidence of Alfv\'{e}nic wave activity \mody during neutral beam heating \norm in KSTAR plasmas. 
In 2010 and 2011 campaigns KSTAR plasmas included 1.2~MW of 
neutral beam heating, which provided a source of heating to excite Alfv\'{e}nic wave activity modes. 
Data from the 2010 campaign, which was fully analysed during 2011, shows 40-60kHz magnetic fluctuations. 
\mody We present the first ideal MHD calculation of a core localised mode with toroidal mode number $n=1$
for an experimental configuration. The mode is global with very small resonance with continuuum modes.  %: previous ideal MHD BAE treatments exhibit significant resonance with continuum modes.  
Second, this work is the first observation of Alfv\'{e}nic wave activity in KSTAR. 
With up to 14MW of neutral beam heating and 14MW of RF heating planned, KSTAR plasmas will become a pilot for ITER plasmas, and provide the opportunity to explore 
the wave-particle-plasma interaction in regimes approaching burning plasmas. Our work builds on preliminary observations of electron fishbones in KSTAR in 2009, \cite{Ryu_10} and 
aliased TAE activity in 2011. \cite{Hole_IAEA_2012} Finally, as a spin-off, we have developed a new form of MHD spectroscopy for consistency of the temperature profile with the observation of wave activity. 
The remainder of the work is as follows: \norm
in Section 2 we introduce the experiments conducted in KSTAR, and in Sec. 3 present detailed modelling. Finally, Sec. 4 contains concluding remarks and discusses implications for future work.

% #################################################################################################
\section{Experiments}

In 2010 a set of neutral beam injection (NBI) excitation experiments were conducted in KSTAR in an attempt to generate shear Alfv\'{e}nic wave activity.
Subject to operator controls, the choice of plasma conditions were optimised for this purpose:
a relatively low toroidal magnetic field of 1.95~T, and maximum available NBI heating: 1.2MW of 80keV NBI.
For deuterium plasmas, 80keV NBI would produce fast D neutrals with speed $v_{||,\mathrm{beam}} = 2.8 \times 10^6$~ms$^{-1}$. 
With plasma densities of up to  $ 5 \times 10^{19}$~m$^{-3}$ expected, the minimum Alfv\'{e}n speed is $v_A = 4.4 \times 10^6$~ms$^{-1}$.
While the beam speed is sub-Alfv\'{e}nic, it is greater than the first sideband resonance at $v_A/3$, and so it is possible that TAE 
modes may excited, as was found during early operation of MAST. \cite{Appel_02} Alfv\'{e}n cascades have also been observed on JET in reverse shear configurations with
$v_{||,\mathrm{beam}}/v_A$ as low as 0.2. \cite{Sharapov_06}

\begin{figure}[h]
\centering
\includegraphics[width=80mm]{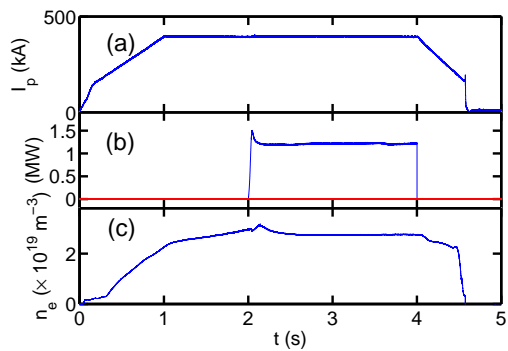}
\caption{\label{fig:4220_evolution} 
Evolution of \#4220 showing (a) plasma current $I_p$, (b) auxiliary heating ($P_{NBI}$ in blue, $P_{ECRH}$ in red), and
(c) line averaged electron density $n_e$.}
\end{figure}

Four NBI heated discharges (\#4218-\#4220) were produced, with flat top plasma current $I_p$ in the range $210< I_p< 407$~kA, electron cyclotron resonance heating (ECRH) power up to 200~kW, and 
core plasma density up to $ 3 \times 10^{19}$~m$^{-3}$. There was no discernible impact of ECRH heating on these plasmas. 
Figure \ref{fig:4220_evolution} shows the temporal evolution of one such discharge, \#4220.
For this 4.5s plasma,  2s of NBI was applied from 2s during current flat-top. 
% In Fig. \ref{fig:4220_evolution}(d) inferred toroidal plasma beta is also shown, inferred from EFIT calculations without MSE or pressure profile constraints. 

\begin{figure}[h]
\centering
\includegraphics[width=80mm]{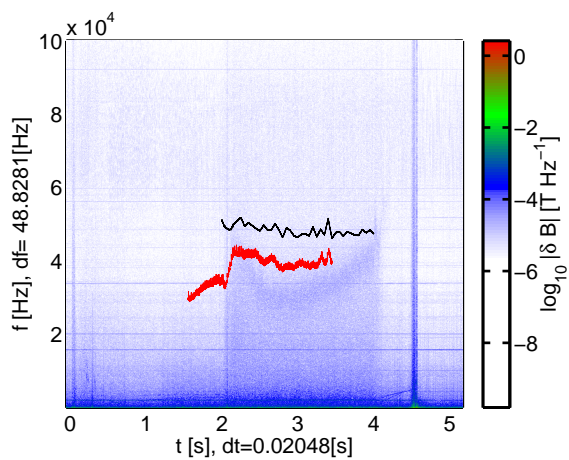}
\caption{\label{fig:4220_spectrum} 
Magnetic spectrogram of \#4220 with $ \omega_{CAP}$ (red) and BAE mode (black) from  Fig. \ref{fig:4220_mishka}(a) overlaid (red).}
\end{figure}

Figure \ref{fig:4220_spectrum} shows magnetic oscillations of discharge \#4220. 
The mode activity correlated with NBI heating, and a study of electron cyclotron emission data reveals that the plasma is sawtoothing throughout the heating phase. 
A study of signal phase versus geometric angle, computed from a toroidal Mirnov array provides weak evidence that the mode has $n=1$.  %\mody and that the mode propagates in the ion diamagnetic direction. \norm  
Unfortunately, the coils are located flush with the conducting wall, and so the signal to noise ratio is large, and there is significant $n=0$ noise present.
Information about the poloidal mode number is not available. Similar oscillations were observed in discharges \#4218,\#4219 and \#4221. 

A simple scoping exercise reveals that the mode activity is unlikely to correspond to a Toroidal Alfv\'{e}n Eigenmode. 
The middle of the TAE gap lies at $\omega_{TAE}=v_A/(2 q R)$. Using $n_i \approx n_e(0)$, taking $R$ as the magnetic axis, and using
$q=q_{mn}=(2m+1)/(2n)$ for $m=1,2,3$ gives 160~kHz, 100~kHz and 70~kHz, respectively. 
\modb Mode activity with frequency of order 150~kHz and $m=n=1$ was observed in the 2011 campaign. \norm 
The observed mode \modb reported here \norm has a frequency of 40~kHz, which would be commensurate with 
a resonance of $q=5$. This is the edge $q$ of these elongated plasmas, and so the TAE frequency will be significantly greater than 40~kHz. 
\modb Following Gorelenkov \etal \cite{Gorelenkov_09} we have also computed the thermal ion transit frequency $\omega_{ti} = \left . {\sqrt{2 k_B T_i/m_i}} \right / (q R_0)$,
which is the upper beta-acoustic Alfv\'{e}n Eigenmode (BAAE) gap frequency.   For KSTAR plasmas, the on-axis frequency ranges from 29kHz, using the central 
ion temperature inferred from \#4229 down to 17kHz. These are below the observed wave frequency, at 40-60~kHz. \norm

In contrast, the frequency as well as its time evolution is a closer match to the evolution of the kinetic accumulation frequency 
$ \omega_{CAP}= 1/R_0 \sqrt{2 T_i/m_i (7/4 + T_e/T_i))}$, with $R_0$ the major radius, $m_i$ the ion mass, and $T_i$ and $T_e$ the ion and electron temperature, respectively. \cite{Zonca_96} 
Electron cyclotron emission (ECE) data gives the on-axis value of electron temperature $T_e = 1.2$~keV.
While not available for discharge \#4220, x-ray imaging crystal spectrometer \cite{Lee_XECS_10} data providing $T_i, T_e$ is available for nearby discharge \#4229 
with the same level of NBI heating. \mody This pulse also shows evidence of toroidal rotation, with core rotation up to 100~km/s, 
producing a core Doppler shift of up to 8kHz. Rotation of either a near stationary mode or magnetic island is thus insufficient to describe the observations.\norm 

Correcting for the offset in neutral beam heating interval for this discharge, we have computed $\omega_{CAP}$ and over plotted the evolution in Fig. \ref{fig:4220_spectrum}.   
The frequency match is close, as is the slight frequency drop following beam turn-on at 2~s. The drop in frequency occurs due to the initial drop
in $T_e$ observed in the core ECE channel. Similar BAE frequency scaling is evident for discharges \#4218 and \#4219. 

% #################################################################################################
\section{Detailed Modelling} 

\begin{figure}[h]
\centering
\includegraphics[width=80mm]{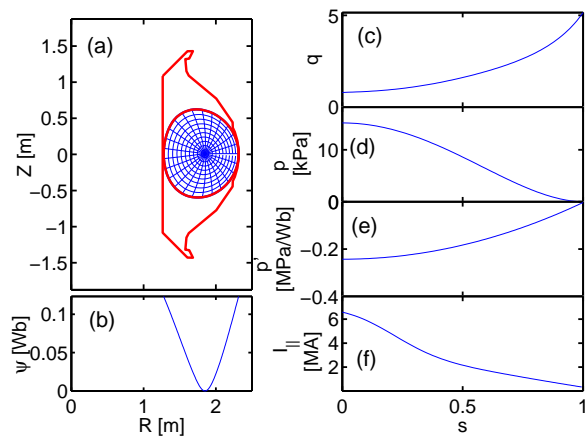}
\caption{\label{fig:4220_equilibrium} 
Equilibrium for \#4220 at 2s. Panel (a) shows contours of poloidal flux with the plasma vessel cross section  overlaid, and panel (b)
is a major radius profile of poloidal flux $\psi$. Panels (c)-(f) show $q, p, p'(\psi)$ and $I_{||}$ as a function of $s$, the square root of normalised poloidal flux. }
\end{figure}

We have undertaken detailed modelling of the plasma at the onset of mode activity at 2~s. 
By analysing data from a set of 20 electron cyclotron emission (ECE) chords we have been able to identify the inversion radius and locate the $q=1$ surface.  
To correct the $q=1$ surface of magnetics-only constrained EFIT to match ECE data 
we have used CHEASE \cite{CHEASE_96} to remap the current profile as $I^*(s) \rightarrow I^*(s)+ c I_{core}(s)$
and pressure profile with $p'(s) \rightarrow \lambda p'(s)$ so as to match the sawtooth inversion radius and EFIT $\beta_p$, respectively.  
Here, $s$ is the square root of normalised poloidal flux, with $s=0$ the core and $s=1$ the edge.
The core current profile selected is a ramp, $I_{core}(s) =1-s.$  
Figure \ref{fig:4220_equilibrium} shows a cross section of the corrected equilibrium flux surfaces and $q$ profile.  

\begin{figure}[h]
\centering
\includegraphics[width=60mm]{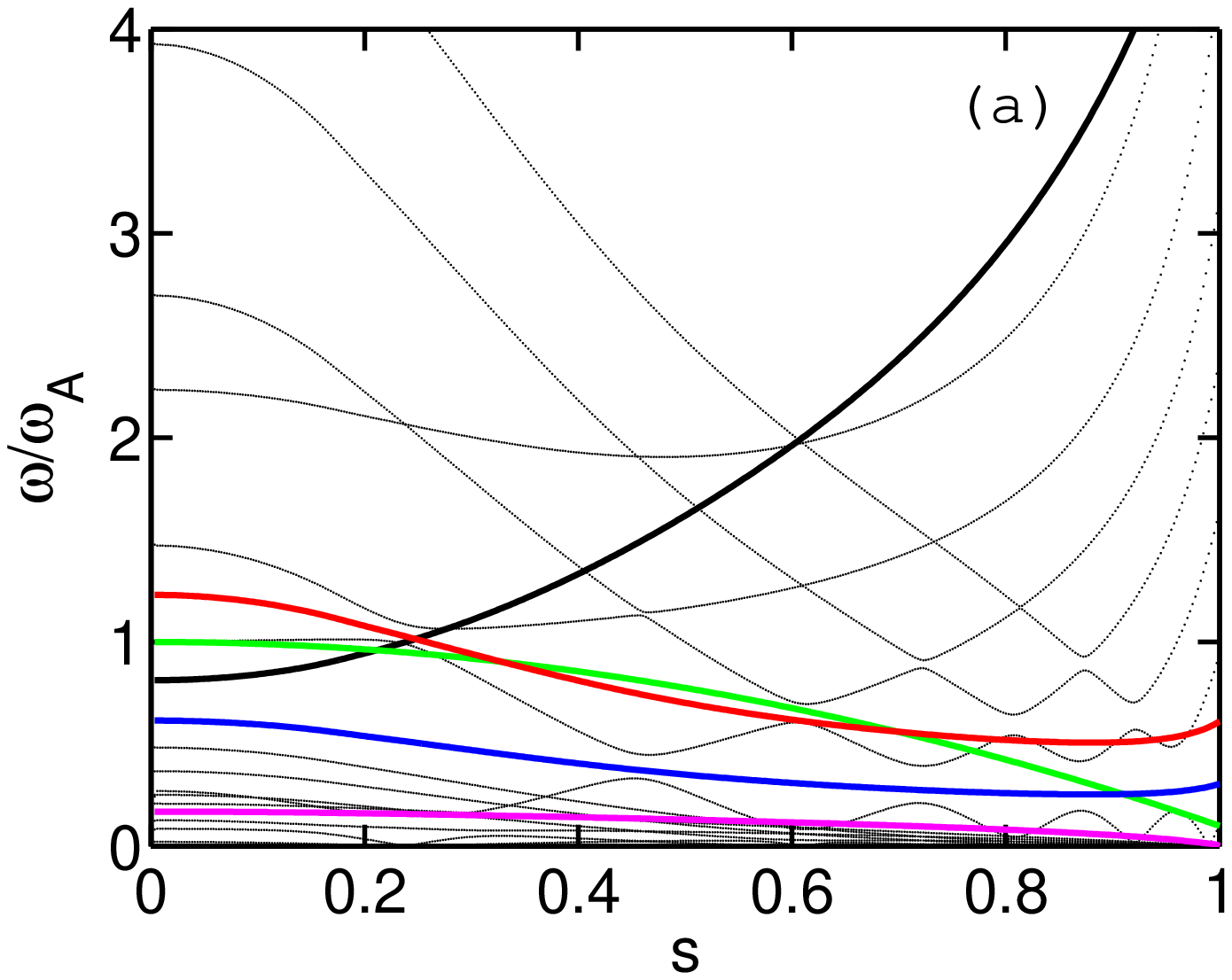}
\includegraphics[width=60mm]{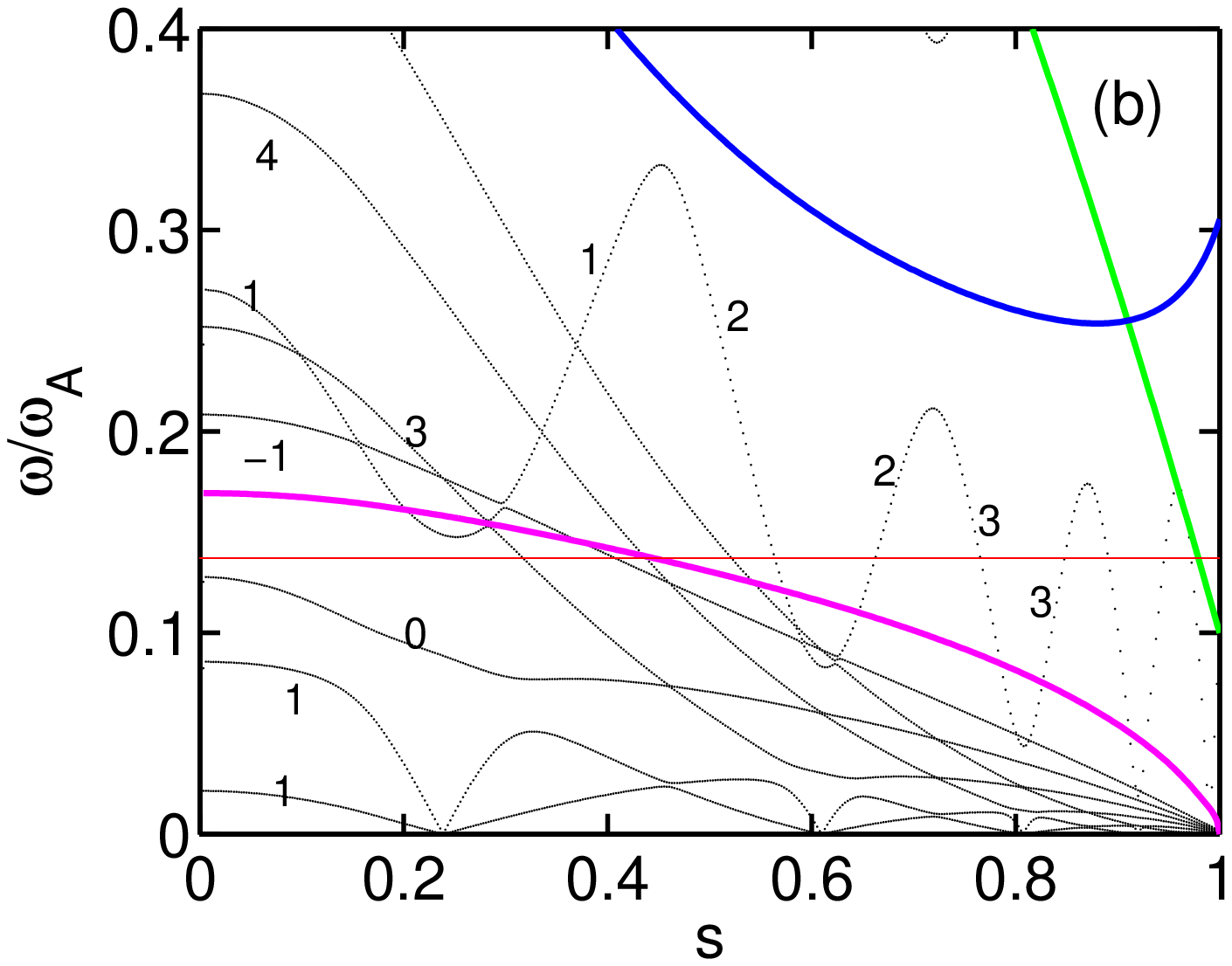}
\caption{\label{fig:4220_cscas} 
Ion sound and Shear Alfv\'{e}n continuum (shown as black points) for $n=1$ modes for \#4220 at 2s. 
Also shown is the $q$ profile (solid black curve), normalised mass density profile $\rho/\rho_0$ (solid green curve), 
TAE frequency (solid blue curve), elliptical Alfv\'{e}n eigenmode frequency (solid red curve) and accumulation point of the low frequency gap $\omega_{As,gap}$ (solid pink curve).
Figure (b) focuses on the BAE region, which also shows the BAE mode (red line).}
%
% the upper and lower accumulation points (red lines) for the restricted domain $0<s<0.3$ together with the BAE mode (red dashed line).}
\end{figure}

Figure \ref{fig:4220_cscas} shows the Alfv\'{e}n and ion sound continuum for $n=1$, computed using the code CSCAS \cite{CSCAS_Poedts_93} with adiabatic index $\gamma=5/3$. 
In Fig. \ref{fig:4220_cscas}(a) the toroidicity and ellipticity induced gaps can be identified.
Using the most recent version of the ideal MHD global stability code MISHKA \cite{MISHKAF_06} we have computed TAE gap modes. 
The TAE gap mode produced by the $q_{mn}=(2m+1)/2n = 1.5$ resonance at $s=0.45$ has a frequency of 120~kHz, well above the measured 40~kHz oscillation.  
In Fig. \ref{fig:4220_cscas}(b) we have zoomed into the low frequency  part of the continuum and identified different continuum branches.
\mody As shown by Gorelenkov \etal, \cite{Gorelenkov_09} the frequency of the ion sound and modified shear Alfv\'{e}n continuum modes drops to zero at rational surfaces. \norm
The accumulation point of the low frequency gap introduced in the shear-Alfv\'{e}n continuous spectrum because of finite beta is $\omega_{As,gap}/\omega_A = (\gamma \beta)^{0.5}$,
with $\omega_A$ the Alfv\'{e}n angular frequency at the magnetic axis.
To identify a global mode separated from the MHD continuum, we have selected an $s$ interval of $0<s<0.3$, thereby avoiding the resonance with the $m=3$ ion sound branch. 
The discrete mode identified in Fig. \ref{fig:4220_cscas}(b) is separated from the continuum above and below, and so forms a gap mode between singular modes. 
In the limit that the $s$ interval is expanded to the full domain, the eigenfunction retains its global structure. 
\mody The mode frequency is $\omega/\omega_A = -0.1371$. A single channel millimeter wave interferometer system provides a measurement of the time-resolved line integrated
electron density, $n_e$.\cite{Nam_08} Assuming a parabolic profile for the density provides a measure of on-axis density. If $n_i \approx 0.8 n_e$ is also assumed, \cite{Hole_05} 
the Alfv\'{e}n angular frequency can be computed, yielding $\omega/(2 \pi) \approx 50$~kHz at 2s. We have overplotted the frequency evolution of this
estimate in Fig. \ref{fig:4220_spectrum}. \norm

\begin{figure}[h]
\centering
\includegraphics[width=60mm]{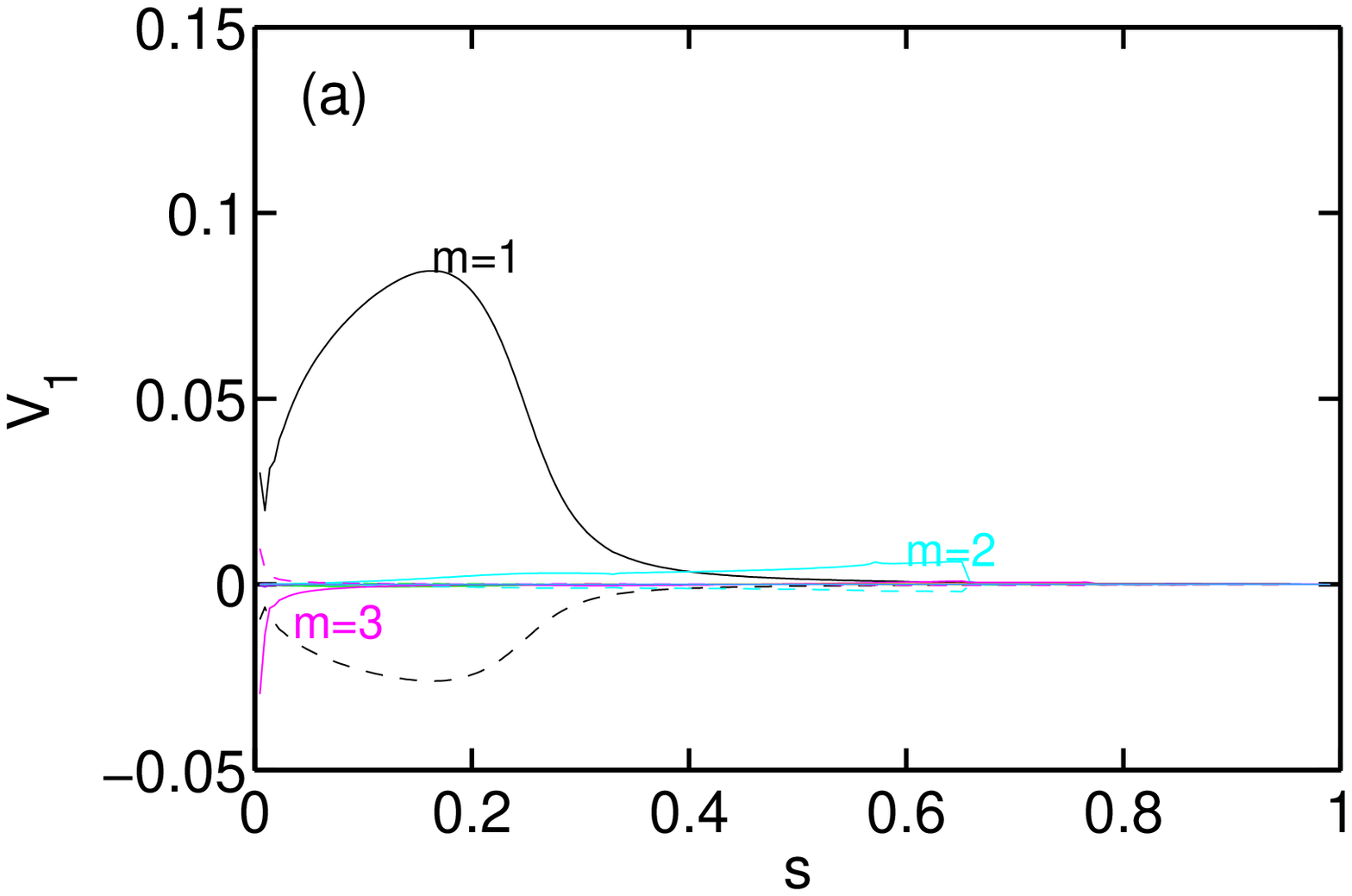}
\includegraphics[width=60mm]{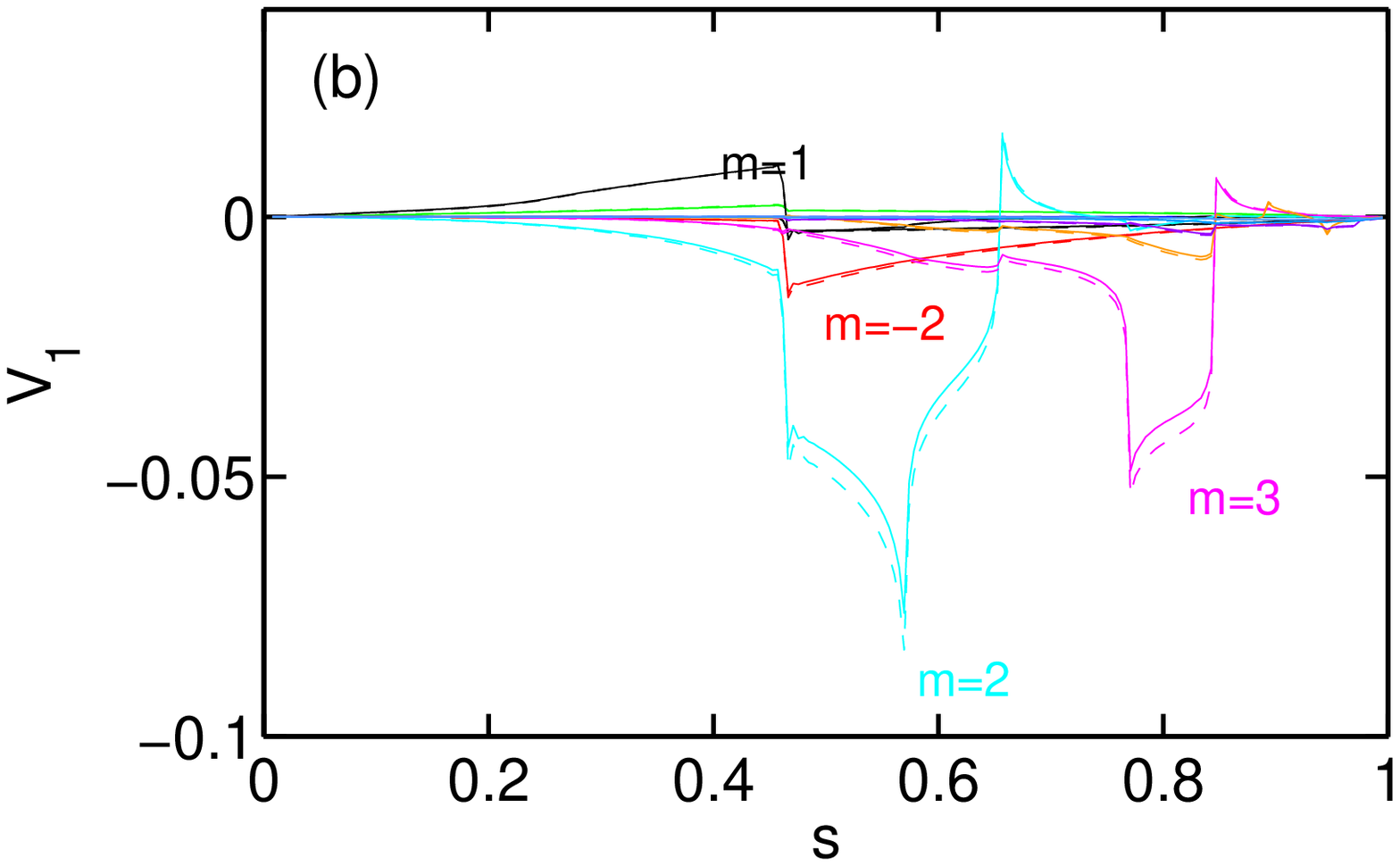}
\caption{\label{fig:4220_mishka} 
Eigenfunctions for (a) an $n=1$ BAE with frequency $\omega/\omega_A = -0.1371$, and (b) an $n=1$ continuum mode with frequency $\omega/\omega_A = -0.1380$.
Both real (solid) and imaginary (dashed) components are shown.}
\end{figure}

The mode, which is resonant with the core of the plasma, has similar global mode structure to modes computed in DIII-D by Turnbull \etal \cite{Turnbull_93}.
In contrast, the BAE mode modelled by Huysmans \etal \cite{Huysmans_95} is in the region of greater shear, and posseses more poloidal harmonics. 
Figure \ref{fig:4220_mishka} shows the $V_1$ eigenvector, which is the $\rho$ component of the contravariant fluid displacement velocity, 
for (a) a discrete mode, and (b) a nearby continuum mode. 
\mody A continuum mode is one that is localised to a resonant surface with poloidal flux $\psi$, and a solution of the 
shear Alfv\'{e}n dispersion relation $\omega(\psi) = k_{||}(\psi) v_A(\psi)$, with $k_{||}=\bk \cdot \bB/B$.\cite{CSCAS_Poedts_93} 
The BAE in Fig. \ref{fig:4220_mishka}(a) has a global mode structure with negligent resonant coupling with continuum modes, whereas
the continuum mode in  Fig. \ref{fig:4220_mishka}(b) has  resonances at crossings of the continuum at $s=0.57, s=0.66$ for $m=2$ and $s=0.77, s=0.84$ for $m=3$. 
In contrast to earlier work, \cite{Turnbull_93, Huysmans_95} the global mode of Fig. \ref{fig:4220_mishka}(a) has significantly reduced coupling to 
resonances with the continuum, and therefore will exhibit weaker continuum damping. \norm

\mody Ideal MHD codes such as MISHKA solve for the wave structure in full toroidal geometry, but provide no information about mode drive. 
% An esimate of whether the fundamental or sidebands can be made 
% Calculation of the resonance condition, together with information about the beam velocity can determine whether the fundamental or sidebands cmode is driven. 
\modc  The particle-wave resonance condition \cite{Heidbrink_08} is
\begin{equation}
\frac{n}{R} v_{||} - \frac{m+l}{q R} v_{||} - \omega = 0, \label{eq:resonance}
\end{equation}
where $l$ is a Fourier mode number in poloidal angle $\theta$ of the particle magnetic drift velocity. \norm
The mode peaks at $s=0.16$, for which $q=0.89$. 
At this radial location \modc the $l=0$ resonance condition requires the unphysical condition $v_{||}/v_{||,beam}=1.3$, 
while for $|l|\ge 1$ the resonance condition can be satisfied for $v_{||} < v_{||,beam}$. 
Thus, the mode can be driven by sidebands, $|l|\ge 1$.
The mode resonance is broad due to the finite radial region of varying $q$ covered by the mode width.

Meanwhile, the mode excitation threshold by energetic ions is reduced by
inverse ion Landau damping connected with finite thermal ion temperature
gradient. \cite{Zonca_96} \norm
\modb In the limit of vanishing continuum damping, \norm an estimate of the threshold to marginal stability can be made using the kinetic treatment of Zonca \cite{Zonca_96}, which 
\mody studied the drive due to thermal ion temperature radial gradients, and \norm 
studied the relationship between kinetic ballooning modes (KBM) and BAEs.  
In that treatment it was shown the stability conclusions were a function of 
the frequency range of $\omega_{CAP}$ relative to the core-plasma ion diamagnetic frequency
$\omega_{*pi} = \frac{k_B T_i}{e_i B} (\bk \times \bb) \cdot \nabla \ln P_i$. \norm
Here, $k_B$ is Boltzman's constant, $\bk = m/r \mathbf{e_\theta} + n/R \mathbf{e_\phi}$ the wave vector, $\bb = \bB/|\bB|$, \mody and $P_i$ the thermal ion pressure.
If $\omega_{*pi}^2 \ll \omega_{CAP}^2$ the KBM accumulation point was always stable, and the 
\norm BAE may become unstable for values of $\eta_i = (\partial \ln T_i / \partial \ln n_i)$ greater than a critical value $\eta_{ic}$, 
given by 
\begin{equation}
\eta_{ic} \approx \frac{2}{\sqrt{7 + 4 \tau}} \frac{\omega_{ti}}{q \omega_{*ni}}, \label{eq:eta_ic}
\end{equation}
with % $\omega_{ti} = \left . {\sqrt{2 k_B T_i/m_i}} \right / (q R_0)$ the thermal ion transit frequency, and 
$\omega_{*ni} = \frac{k_B T_i}{e_i B} (\bk \times \bb) \cdot \nabla n_i / n_i$ the ion diamagnetic drift frequency. 
\mody If $\omega_{*pi}$ is increased further, the unstable BAE accumulation point is expected to smoothly connect to 
an unstable KBM accumulation point with exponentially small growth rate when $\omega_{*pi}^2 \gg \omega_{CAP}$. 
The most unstable BAE/KBM accumulation point occurs when $\omega_{*pi}^2 \approx \omega_{CAP}^2$, when the BAE and KBM are strongly coupled. 
For the plasma conditions of KSTAR, we compute $\omega_{*pi}/(2 \pi) \approx 4~$kHz, and so $\omega_{*pi}^2 \ll \omega_{CAP}^2$, and thus 
\modb plasmas with $\eta > \eta_{ic}$ with $\eta_{ic}$ given by Eq. (\ref{eq:eta_ic}) are unstable. \norm
In this frequency regime $\delta E_\|\approx 0$, \modb consistent with ideal MHD. \norm 
Indeed, we compute $\omega_{As, gap}/(2 \pi) \approx 60$~kHz at the minimum of the $m=1$ continuum at $s=0.24$.  
This frequency, and the computed BAE gap mode frequency of 50~kHz at 2s, is close to the kinetic accumulation frequency $\omega_{CAP}$. \norm

By expanding $\eta_i = (\partial \ln T_i / \partial \ln n_i) = \left . { \frac{\partial \ln T_i}{\partial r}} \right /  \frac{\partial \ln n_i}{\partial r}$, 
as well as $(\bk \times \bb) \cdot \nabla n_i / n_i = (\bk \times \bb)_r  \frac{\partial \ln n_i}{\partial r}$, the ratio $\eta_i/\eta_{ic}$ expands as
\begin{equation}
\eta_i / \eta_{ic} = \frac{\partial \ln T_i}{\partial r} \frac{k_B T_i/(e_i B) (\bk \times \bB)_r}{\sqrt{2/(7 + 4 \tau)} \omega_{*i}/q} \label{eq:eta_ratio}
\end{equation}
Equation (\ref{eq:eta_ratio}) is independent of the ion density profile, and so can be computed for different radial temperature profiles.

X-ray crystallography of nearby discharge \#4229 measures $T_i(0)/T_e(0) = 0.75$ during NBI heating, and we have used this value for \#4220.
Together with  ECE $T_e$ data for \#4220, this fixes $T_i$ on-axis. 
To account for the unknown core localisation of the ion temperature profile, we have expressed the temperature profile
as the sum of profiles $T_a(s)$ and $T_b(s)$, with $T_i(s) = T_a(s) + T_b(s)$, and $r \approx s a$, with $a$ the minor radius.
The profile $T_a(s)$ is an approximation to the Ohmic temperature profile.
A constraint for $T_a(0)$ is provided by crystallography during the pre-NBI heating phase of \#4229: that is $T_a(0)= 0.25 T_e(0)$.
As a plausible estimate for the Ohmic ion temperature profile we have assumed $n(s) \propto T_a(s)$, and inferred $T_a(s)$ from the equilibrium pressure profile. 
Finally, we have modelled the core temperature profile as
\begin{equation}
T_b(s)=(T_i(0) - T_a(0))(1-\tanh(\alpha s)) \tanh(10 (1-s)), \label{eq:Tb_profile}
\end{equation}
with $\alpha$ varied to control the core radial localisation of $T_b(s)$, and term $\tanh(10 (1-s))$ included to force $T_b(s)$ to zero at $s=1$.  

\begin{figure}[h]
\centering
\includegraphics[width=80mm]{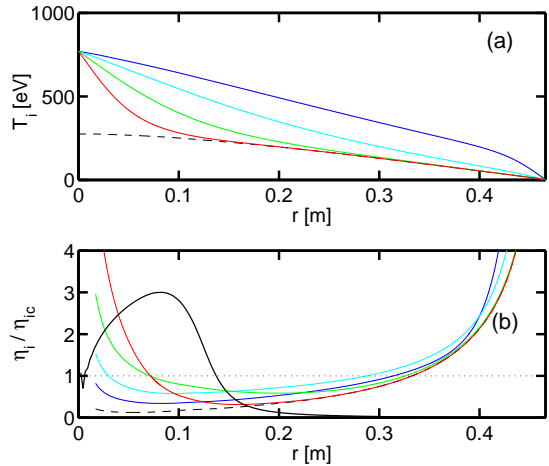}
\caption{\label{fig:4220_threshold} 
Variation of $\eta_i/\eta_{ic}$ threshold with different ion temperature profiles. Panel (a) shows possible ion temperature profiles (solid) with core temperature matching 
crystallography data from discharge \#4229, and panel (b) shows the corresponding profile of $\eta_i/\eta_{ic}$ . In both panels the dashed line corresponds
to a possible ohmic ion temperature profile, with $\tau$ taken from discharge \#4229 prior to NBI heating. 
In panel (b) the light line is $|V_1|$ of the BAE in Fig. \ref{fig:4220_mishka}(a).}
\end{figure}

Figure \ref{fig:4220_threshold}(a) shows a plot of different candidate modelled ion temperature profiles on the outboard radial chord, and Fig. \ref{fig:4220_threshold}(b) shows the corresponding 
$\eta_i / \eta_{ic} $ values. The candidate BAE mode in Fig. \ref{fig:4220_mishka}(a), 
\modb whose $m=1$ Fourier magnitude is also shown in Fig. \ref{fig:4220_threshold}(b), \norm 
has a peak at radial position $r \approx 0.08$~m with a radial width of $0.2a=0.1$~m.
For this mode,  $\eta_i/\eta_{ic}<1$ for a broad, Ohmic-like ion temperature profile.
\modc For sufficiently high radial
temperature gradient and/or sufficiently high ion temperature, the Alfv\'{e}nic
ion temperature gradient driven mode \cite{Zonca_96} instability threshold will be
approached, or possibly even exceeded, in the region where the mode
amplitude is large, and so the mode can become unstable, due to a
combination of energetic and thermal ion kinetic effects \cite{Zonca_99}.\norm

% \mody Indeed, the plasma will be unstable to BAE modes for all but the very broadest of ion temperature profiles, which has a pedestal and would represent H-mode, and KSTAR did not access H-regime plasmas in 2010. 
% Thus, these observations constitute some experimental validation of the technique. \norm
% The inference of temperature profiles in the radial interval of the mode from mode existence is complementary to recent work by Bertram \etal, who
% suggested using mode frequency to infer the temperature at the resonant surface of a BAE mode in the H-1 heliac. \cite{Bertram_12} 

% #################################################################################################
\section{Conclusions}

We have provided first evidence of beta induced Alfv\'{e}n eigenmode activity in neutral beam heated KSTAR plasmas.
The 40~kHz, $n=1$ observed mode matches the frequency of the accumulation point of the Alfv\'{e}n continuum.
By using the radial localisation of sawtooth inversion radius, we have been able to identify the radial position 
of the $q=1$ surface, and using this, constrain the equilibrium. A detailed mode analysis reveals the presence 
of a core localised beta induced Alfv\'{e}n eigenmode. Finally, a kinetic treatment of the mode marginal stability threshold
shows a range of plausible ion temperature profiles for which the mode 
\modc exctiation threshold is reduced, and the mode can be driven by energetic ions. \norm
This suggests that mode existence could be used as a form of inference for 
temperature profile consistency in the radial interval of the mode, 
\modc once beam drive and relevant damping contributions have been computed, \norm
thereby extending the tools of MHD spectroscopy.
Our suggestion is complementary to recent work by Bertram \etal,\cite{Bertram_12}  who
suggested mode frequency could be used to infer the temperature at the resonant surface of a BAE mode in the H-1 heliac. 

Improvements in diagnosis and reliability of KSTAR plasmas will enable further exploration of new physics, in steady state plasma
environments of up to 14MW of NBI heating. A priority is the introduction of kinetic and motional Stark effect constraints to EFIT, which will remove uncertainties in detailed modelling. 
In future work we hope to consolidate mode frequency variation with pressure and magnetic field strength. 
Judicious phasing of NBI during current ramp-up and ramp-down, together with new electron cyclotron current drive systems
also offer the opportunity to influence the magnetic field configuration through co and counter injection. A wider sample set 
of modes, together with concurrent crystallography data, would enable a more thorough investigation of marginal stability thresholds
\modc and a quantitative study of mode drive and damping.  \norm \modb
Finally,  as the position of the resonances are determined by the $q$ profile and the density profile, similarity experiments with
different edge $q$ profiles and density profiles would illuminate the level of continuum damping. 
For instance, as the $q$ profile flattens different $m$ continuum mode branches are removed, and continuum damping is reduced. 
Alternately, as the density profile in the edge drops the location of the continuum resonance moves inward. 
In this case the radial separation between the resonances and gap mode decreases, 
the coupling of the eigenfunction to the continuum becomes stronger, and hence the continuum damping is increased. \cite{Berk_damping_92,Zonca_92}   
\norm

% #################################################################################################

\section*{Acknowledgments}
This work was partly funded by the Australian Government through Australian Research Council grants FT0991899, DP1093797, 
as well as National Research Foundation of Korea grants NFR 2011-0018742 and NRAF-2012-0000590, and the RCUK Energy Programme under
grant EP/I501045 and the European Communities under the contract of
Association between EURATOM and CCFE. The views and opinions expressed
herein do not necessarily reflect those of the European Commission.
The authors gratefully acknowledge the support of J. Kim from NFRI in assisting with access to KSTAR data, and 
the support of G. A. Huysmans from CEA Cadarache in the provision of CSCAS, HELENA and MISHKA codes.

\section*{References}
 
\bibliographystyle{unsrt}
\bibliography{MHD,bayesian,varenna_2010,MHD_energetic_particles}

\end{document}